\documentclass[manuscript]{aastex}

\newcommand{\ms}{\noalign{\vspace{3pt plus2pt minus1pt}}}

\newcommand{\be}{\begin{equation}}
\newcommand{\ee}{\end{equation}}
\newcommand{\bea}{\begin{eqnarray}}
\newcommand{\eea}{\end{eqnarray}}

\newcommand{\gapprox}{\lower.4ex\hbox{$\;\buildrel >\over{\scriptstyle\sim}\;$}}
\newcommand{\lapprox}{\lower.4ex\hbox{$\;\buildrel <\over{\scriptstyle\sim}\;$}}

\def\rmd{d}
\def\rme{e}
\def\rmi{i}
\def\bi{\bf}

\shorttitle{Linear acceleration emission 2}
\shortauthors{Melrose and Luo}

\begin{document}

\title{Linear acceleration emission: 2~Power spectrum}
\author{D. B. Melrose and Q. Luo}

\affil{Sydney Institute of Astronomy, School of Physics, University of Sydney, NSW 2006, Australia}

\begin{abstract}
The theory of linear acceleration emission is developed for a large amplitude electrostatic wave in
which all particles become highly relativistic in much less than a wave period. An Airy integral
approximation is shown to apply near the phases where the electric field passes through zero and the
Lorentz factors of all particles have their maxima. The emissivity is derived for an individual
particle and is integrated over frequency and solid angle to find the power radiated per particle.
The result is different from that implied by the generalized Larmor formula which, we argue, is not valid in this case. We also discuss a mathematical inconsistency that arises when one evaluates the power spectrum by integrating the emissivity over solid angle. The correct power spectrum increases as the 4/3rd power of the frequency at low frequencies, and falls off exponentially above a characteristic frequency. 

We discuss application of linear acceleration emission to the emission of high frequency photons in an oscillating model for pulsars. We conclude that it cannot account for gamma-ray emission, but can play a role  in secondary pair creation.
\end{abstract}


\section{Introduction}

The most familiar emission process for highly relativistic electrons (and positrons) is synchrotron emission. Historically, it is of interest that \cite{s49}, in his original development of the theory of synchrotron radiation, also discussed linear acceleration emission (LAE) by relativistic particles. \cite{s49} did not develop the theory for LAE in the same detail as that for synchrotron emission, and there appears to be no subsequent detailed development of the theory of LAE for highly relativistic particles. Our objective in this paper is to develop the theory of LAE for motion in a large-amplitude electrostatic wave (LAEW), emphasizing the analogy with synchrotron emission, and the important differences from synchrotron emission. Two other emission processes for highly relativistic particles, inverse Compton emission and emission due to motion in a large-amplitude transverse wave \citep{go71,a72}, have properties that are somewhat analogous to those of synchrotron emission. In particular, all three are treated by making an Airy-integral approximation to a relevant phase integral. Our treatment of LAE in a LAEW is based on the assumption that an Airy-integral approximation is also appropriate in this case. Our argument for this assumption is given in an accompanying paper \citep{mrl09}, hereinafter referred to as paper~1. We should emphasize that any treatment of LAE encounters conceptual difficulties that are not relevant to the other three emission processes mentioned. In particular, \cite{s49} showed how the theory of synchrotron emission reproduces the total power radiated as predicted by the generalized Larmor formula, and it can be shown that this is also the case for the other two mechanisms mentioned. However, it is not the case for LAE. Moreover, it has been recognized for over a century that there is a conceptual difficulty in the treatment of linear acceleration emission itself, and the underlying difficulty leads to problems in any treatment of LAE.

The motivation for this investigation relates to possible emission processes that can occur in an oscillating model for a pulsar \citep{letal05} or magnetar \citep{bt07} magnetosphere. Specifically, we pose the question whether LAE in an oscillating model can be important as an emission mechanism, and whether the properties of LAE can lead to observational signatures that are unique to an oscillating model. LAE is of potential interest in four ways. First, it may be relevant as a high-energy emission process, which would require that LAE allow emission up to gamma-ray energies. Second, LAE may be relevant to secondary pair creation, which requires photon energies of at least an MeV. Third, all particles in the LAEW emit LAE, and the associated damping of the LAEW is of potential interest in itself, providing a simple way of relating the power in LAE to the energy in the LAEWs. Finally, there is the possibility of a maser form of LAE as a radio emission mechanism, as discussed briefly in paper~1. 

We use the theory in paper~1 with two notable changes. First, in paper~1 we used primes to denote quantities in the primed frame in which the oscillations are purely temporal; the primed quantities are related to those in the laboratory frame, in which the LAEW have a phase speed $v_\phi$ by a Lorentz transformation with velocity $c^2/v_\phi$. In this paper our analysis is restricted to the frame in which the oscillations are purely temporal, and for convenience in writing we omit the primes on all relevant quantities. Second, in paper~1 we concentrated on a triangular wave form, which is an excellent approximation for a LAEW in which the electrons and positrons become highly relativistic, and here we generalize to an arbitrary wave form. This allows us to apply our results for LAE more generally. We described by the wave form by a periodic function, $T(\chi)$, of phase $\chi$, and derive the orbit of a particle by expanding in inverse powers of its Lorentz factor. In Sec.~\ref{sect:emissivity} we derive the emissivity for LAE, and in Sec.~\ref{sect:Airy} we evaluate it in terms of Airy integrals. In Sec.~\ref{sect:Larmor} we compare our results with the generalized Larmor formula and we discuss inconsistencies that arise. We discuss our results and the application to pulsars in Sec.~\ref{sect:discussion} and summarize the conclusions in Sec.~\ref{sect:conclusions}.

\section{Emissivity in LAE}
\label{sect:emissivity}

An exact treatment of LAE is involves assuming period motion and expanding in a Fourier series \citep{r95}. As shown in paper~1, the emission is at harmonics, $\omega=s\Omega$, of the frequency of the LAEW. For a highly relativistic particle, very high harmonics dominate, the sum over $s$ can be replaced by an integral, performed trivially over a $\delta$-function, and the result interpreted in terms of independent pulses of emission each half period of the LAEW. Adopting this viewpoint, we write down the emissivity in LAE  in vacuo for a charge whose orbit is determined by the LAEW.

\subsection{Emissivity}

The energy radiated per unit frequency and per unit solid angle in transverse waves due to a charge, $q=\mp e$ for electrons or positrons, executing an arbitrary one-dimensional motion is given by, e.g., equation (16.18) in \cite{mmcp},
\be
U(\omega,\theta)={\omega^2\over16\pi^3\varepsilon_0c^3}|J(\omega,{\bi k})|^2\sin^2\theta,
\label{Urad1}
\ee
where $\theta$ is the angle of emission with the respect to the axis on the one-dimensional LAEW, with $|{\bi k}|=\omega/c$ for waves in vacuo. From paper~1, the current is given by  
\be
J(\omega,{\bi k})={qc\,\rme^{-\rmi{\bi k}\cdot{\bi x}_0}\over\Omega}
\int\rmd\chi\,\beta(\chi)\rme^{\rmi[\omega\chi/\Omega-k_zz(\chi)]},
\label{LAE10}
\ee
with $\beta(\chi)c$ the velocity and $z(\chi)$ the displacement of the charge as a function of the phase, $\chi$, of the LAEW. 

In the approximation adopted here, there is one pulse of radiation each half period from each particle, with the pulses being in the forward and backward directions in alternative half periods. This allows one to write down the emissivity $\eta(\omega,\theta)$, which is the power radiated per unit frequency and per unit solid angle, by evaluating (\ref{Urad1}) for each pulse, and dividing by the period $2\pi/\Omega$ of the LAEW. This gives
\be
\eta(\omega,\theta)={q^2\omega^2\theta^2\over32\pi^4\varepsilon_0c\Omega}
\left|\int\rmd\chi\,\beta(\chi)\,\rme^{\rmi\phi(\chi)}\right|^2,
\qquad
\phi(\chi)={\omega\over\Omega}[\chi-\cos\theta Z(\chi)],
\quad
Z(\chi)=\int_0^\chi\rmd\chi'\,\beta(\chi'),
\label{eta1}
\ee
with the pulse centered on $\cos\theta=1$ in one half period, and on $\cos\theta=-1$ in the other half period,

\subsection{Motion in a LAEW}

Motion in a LAEW is treated in paper~1. Assuming an electric field $E_0(\chi)=E_0T(\chi)$,  the 4-velocity is $u(\chi)=u_0+(\omega_E/\Omega)F(\chi)$, with $u(\chi)=\gamma(\chi)\beta(\chi)$, $u_0=\gamma_0\beta_0$ a constant of integration, and $F(\chi)=\int_0^\chi\rmd\chi'\,T(\chi')$. Here we assume that the particle is highly relativistic, setting $\beta(\chi)=\pm1$, except in evaluating $Z(\chi)$, which appears in the phase in (\ref{eta1}), where we assume $\beta(\chi)=\pm[1-1/2\gamma^2(\chi)]$, with
\be
\gamma(\chi)=[\gamma^2_0+2u_0(\omega_E/\Omega)F(\chi)+(\omega_E/\Omega)^2F^2(\chi)]^{1/2},
\label{LAE5a}
\ee
$\omega_E=qE_0/m$. A second integration gives the orbit of a particle gives
\be
z=z_0+{c\over\Omega}Z(\chi),
\qquad
Z(\chi)=\int\rmd\chi\,\beta(\chi)\approx\pm\left[
\chi-\int\rmd\chi\,{1\over2\gamma^2(\chi)}\right].
\label{TW3}
\ee 

A background particle, $u_0=0$, is instantaneously at rest at phases $\chi=n\pi$ for any integer $n$ and has its maximum Lorentz factor, $\gamma_{\rm max}$, at the phases $\chi=\pi/2+n\pi$, where we assume $\gamma_{\rm max}\approx|u_{\rm max}|\gg1$, $|u_{\rm max}|=(\omega_E/\Omega)|F(\pi/2)|$. A test charge, $u_0\ne0$, has maximum Lorentz factors, $\gamma_\pm$ at the phases $\chi=\pi/2+n\pi$, and comes to rest, for $|u_0|<|u_{\rm max}|$, at two other phases that are of no interest here. 

\subsection{Basis for Airy-integral treatment}

The basic assumption made here is treating LAE is that the integral over phase, $\phi(\chi)$, in (\ref{eta1}) involves an Airy integral. Physically, this means that the current is dominated by contributions around the phase, $\chi=\chi_0$ say, where the Lorentz factor, $\gamma(\chi)$ has a maximum. In a Taylor series expansion of $Z(\chi)$ and of $\phi(\chi)$ the quadratic term in $\chi-\chi_0$ is then zero, leaving a linear term and a cubic terms, beyond which the expansion is truncated. The integral is then in the form of the standard integral representation of an Airy function. 

For an arbitrary wave form, we are free to choose $F(0)=0$ and $T(\pi/2)=0$, and to write $\gamma_{\rm max}=(\omega_E/\Omega)|F(\pi/2)|$. The extrema of $\gamma(\chi)$ are $\gamma_\pm=\gamma_{\rm max}\pm|u_0|$ at $\chi=\pi/2$, where we assume $\gamma_{\rm max}>|u_0|$, and the sign is determined by the sign of $u_0$ and of the charge.  This gives
\be
Z(\chi)-\chi=-{1\over2\gamma_\pm^2}\left[
\chi'
+{\alpha_\pm^3\over3}\chi'^3\right],
\qquad
\alpha_\pm^3=-{\gamma_{\rm max}\over\gamma_\pm}{T'(\pi/2)\over F(\pi/2)}
\label{TW4b}
\ee
with $\chi'=\chi-\pi/2$. For the triangular wave form one has $-T'(\pi/2)/F(\pi/2)={8/\pi^2}$.

For the general case  (\ref{TW4b}), the phase in (\ref{eta1}) becomes
\be
\phi(\chi)\approx
{\omega\over2\Omega\gamma_\pm^2}\left[(1+\theta^2\gamma_\pm^2)\,
\chi'
+{\alpha_\pm^3\over3}\,\chi'^3\right].
\label{phase0}
\ee
The two contributions of the form (\ref{phase0}), centered on $\chi=\pi/2$, $\chi=3\pi/2$, with $\gamma=\gamma_\pm$, involve emission in a small cone about the forward and backward directions, respectively, and these are treated independently. 

\section{Airy integral approximation}
\label{sect:Airy}

Assuming that the important contribution to the current for LAE is dominated by phases centered on that at which $\gamma(\chi)$ has a maximum, it is straightforward to evaluate the emissivity in terms of an Airy function. The integral over frequency and solid angle to find the total power radiated involves only standard integrals. However, the integral over solid angle to find the power spectrum (the power per unit frequency), although seemingly straightforward, leads to inconsistency when performed in two different ways.

\subsection{Airy integral}

The Airy integral is \citep{AS65}
\be
\int_{-\infty}^\infty\rmd\chi\,\rme^{\rmi(a\chi+b^3\chi^3/3)}={2\pi\over b}{\rm Ai}\left({a\over b}\right)
=2\left({a\over3b^3}\right)^{1/2}K_{1/3}(\xi),
\label{Airy1}
\ee
where in the second form, terms of a Macdonald function, follows from
\be
{\rm Ai}(z)={\sqrt{z}\over\pi\sqrt{3}}\,K_{1/3}(\xi),
\qquad
z={a\over b},
\qquad
\xi={2\over3}z^{3/2}.
\label{Airy2}
\ee
For a test particle, using (\ref{phase0}), one needs to add subscripts $\pm$ to $a\to a_\pm$, $b\to b_\pm$ in (\ref{Airy1}), and to $z,\xi$ in (\ref{Airy2}). One has
\be
a_\pm={\omega\over\omega_{c\pm}}(1+\gamma_\pm^2\theta^2),
\qquad
b_\pm^3={\omega\over\omega_{c\pm}}\alpha_\pm^3,
\qquad
z_\pm={a_\pm\over b_\pm},
\qquad
\omega_{c\pm}=2\Omega\gamma_\pm^2.
\label{Airy4}
\ee
For simplicity in writing, in the following formulae we omit the $\pm$ subscripts, and assume $a,b,z,\xi,\omega_c$ are given by (\ref{Airy4}). The particular case of a background particle in a triangular wave form (paper~1), one has $\gamma\to\gamma_{\rm max}$, $\alpha^3\to8/\pi^2$.

The emissivity (\ref{eta1}) may be written either in the form
\be
\eta(\omega,\theta)={q^2\omega_c\over4\pi^2\varepsilon_0c}\,
\Theta\,z_c^2{\rm Ai}^2\big(z_c(1+\Theta)\big),
\qquad
\Theta=\gamma^2\theta^2,
\qquad
z_c=\left({\omega\over\omega_c}\right)^{2/3}{1\over\alpha},
\label{eta2a}
\ee
or in the form
\be
\eta(\omega,\theta)={3q^2\omega_c\over16\pi^4\varepsilon_0c}\,\Theta(1+\Theta)
\xi_c^2K^2_{1/3}\big(\xi_c(1+\Theta)^{3/2}\big),
\qquad
\xi_c={2\over3}\,{\omega\over\omega_c}\,{1\over\alpha^{3/2}}.
\label{eta2}
\ee
For emission in the backward direction is given one formally needs to replace $\theta$ by $\pi-\theta$. 

The emissivity (\ref{eta2}) is analogous to the corresponding formula for synchrotron emission \citep{gs65}, with the cyclotron frequency replaced by the frequency, $\Omega$, of the LAEW. An important difference is that for LAE the angular distribution is confined to a small cone about the direction of motion, being zero strictly along this direction, $\theta=0$, whereas for synchrotron emission the maximum occurs where the angle, $\theta$, of emission is equal to the pitch angle of the particle. The dependence on $\theta$ in (\ref{eta2a}), (\ref{eta2}) causes mathematical difficulties, which we gloss over in the following discussion and address explicitly in Sec.~\ref{dilemma}.

\subsection{Angular distribution of LAE}

The power radiated by an individual particle can be evaluated by integrating the emissivity over frequency and solid angle. Performing the integral over frequency first gives the power radiated per unit solid angle, which described the angular distribution of the emission, and performing the integration over solid angle first leads to the power spectrum. 

The integral over $\omega$ can be rewritten as an integral over $\xi_c$ and evaluated using the identity
\be
\int_0^\infty\rmd\xi\,\xi^2K_\mu^2(\xi)={\pi^2(1-4\mu^2)\over32\cos\pi\mu},
\label{schwinger1}
\ee
which gives $5\pi^2/144$ for $\mu=1/3$. (There is an error by a factor of 2 in the counterpart of (\ref{schwinger1}) given by \cite{s49}.) The integral over frequency  gives the power per unit solid angle:
\be
\int_0^\infty\rmd\omega\,\eta(\omega,\theta)=
{9q^2\omega_c^2\alpha^{3/2}\over64\pi^4\varepsilon_0c}\,
{\Theta\over(1+\Theta)^{7/2}}\int_0^\infty\rmd\xi\,\xi^2K_{1/3}^2(\xi)
={5q^2\omega_c^2\alpha^{3/2}\over1024\pi^2\varepsilon_0c}\,
{\Theta\over(1+\Theta)^{7/2}},
\label{Larmor4}
\ee
with $\Theta=\gamma^2\theta^2$, and where (\ref{schwinger1}) gives $5\pi^2/144$ for $\mu=1/3$. It follows that the emission is zero for $\theta=0$, and is strongly concentrated at angles $\theta\sim1/\gamma_\pm$, $\pi-\theta\sim1/\gamma_\mp^2$, for forward and backward emission.

\subsection{Power spectrum for LAE}

Carrying out the integral over solid angle first leads to an expression for the power per unit frequency in LAE. In the case of synchrotron radiation the integral is well known \citep{gs65}, and was derived for this specific purpose by \cite{w59}. The integral needed in the case of LAE is different from that for synchrotron radiation. To avoid the mathematical difficulty discussed in Sec.~\ref{dilemma}, we integrate the emissivity (\ref{eta2a}) over both frequency and sold angle, and change the variables of integration to $z=z_c(1+\Theta)$ and $z_c$:
\be
P=2\pi\int_0^\infty\rmd\omega\int_0^\infty\rmd\theta\,\theta\,\eta(\omega,\theta)=
{3q^2\Omega\omega_c\over4\pi^2\varepsilon_0c}
\int_0^\infty\rmd z_c\,z_c^{5/2}\int_0^\infty\rmd \Theta\,\Theta\,{\rm Ai}^2\big(z_c(1+\Theta)\big).
\label{eta4}
\ee
By writing ${\rm Ai}^2\big(z_c(1+\Theta)\big)=\int\rmd z\,\delta\big(z-z_c(1+\Theta)\big)\,{\rm Ai}^2(z)$ in (\ref{eta4}), and performing the $\xi_c$-integral over the $\delta$ function, one obtains
\be
P={3q^2\Omega\omega_c\alpha^{3/2}\over4\pi\varepsilon_0c}
\int_0^\infty\rmd z\,z^{5/2}{\rm Ai}^2(z)\int_0^\infty\rmd\Theta\,{\Theta\over(1+\Theta)^{7/2}}.
\label{eta5}
\ee
Alternatively, by performing the $\Theta$-integral over the $\delta$ function, one obtains
\be
P={3q^2\Omega\omega_c\alpha^{3/2}\over4\pi\varepsilon_0c}
\int_0^\infty\rmd z_cz_c^{1/2}\int_{z_c}^\infty\rmd z((z-z_c)\,{\rm Ai}^2(z)
={3q^2\Omega\omega_c\alpha^{3/2}\over4\pi\varepsilon_0c}{4\over15}
\int_0^\infty\rmd z_cz_c^{5/2}{\rm Ai}^2(z_c),
\label{eta5a}
\ee
where the final expression is obtained by partially integrating twice. Noting that the integral over $\Theta$ in (\ref{eta5}) is equal to $4/15$, the two results agree. An alternative form is
\be
P
={3q^2\Omega\omega_c\alpha^{3/2}\over20\pi^3\varepsilon_0c}\int_0^\infty\rmd\,\xi_c\,\xi_c^2K_{1/3}^2(\xi_c).
\label{eta5b}
\ee

The power per unit frequency, $P(\omega)$, is identified by writing the right hand side of (\ref{eta5}), (\ref{eta5a}) or (\ref{eta5b}) as the integral of $P(\omega)$ over $\rmd\omega$. One finds
\be
P(\omega)={q^2\Omega\over60\pi\varepsilon_0c}\left({\omega\over\omega_c}\right)^{4/3}
{\rm Ai}^2(z_c)
={2q^2\Omega\over15\pi^2\varepsilon_0c\alpha^3}\left({\omega\over\omega_c}\right)^2K_{1/3}^2(\xi_c).
\label{eta6}
\ee
The low- and high-frequency limits of (\ref{eta6}) are given by
\be
P(\omega)\approx{2q^2\Omega\over5\pi\varepsilon_0c\alpha^2}
\left({\omega\over\omega_c}\right)^{4/3}
\left\{
\begin{array}{ll}
{\rm Ai}^2(0)
&\quad\omega\ll\omega_c,\\
\ms
\rme^{-2\xi_c}/4\pi z_c^{1/2}
&\quad\omega\gg\omega_c,\\
\end{array}
\right.
\label{eta7}
\ee
with ${\rm Ai}(0)=1/3^{2/3}\Gamma(2/3)=0.355$. Thus, at low frequencies, the power in LAE varies $\propto\omega^{4/3}$. 

\subsection{Total power radiated by a single particle}

The power radiated by a particle in LAE follows from (\ref{eta5b}), with the integral evaluated using (\ref{schwinger1}). The result is
\be
P=
{q^2\Omega^2\gamma^2\alpha^{3/2}\over192\pi\varepsilon_0c}.
\label{Larmor5}
\ee
On restoring the $\pm$ subscripts, (\ref{Larmor5}) gives different results for emission in the forward and backward directions:
\be
P_\pm=
{q^2\Omega^2\gamma_\pm^{3/2}\gamma_{\rm max}^{1/2}\over192\pi\varepsilon_0c}
\left|{T'(\pi/2)\over F(\pi/2)}\right|^{1/2}.
\label{Larmor6}
\ee

\section{Comparison with the Larmor formula}
\label{sect:Larmor}

For synchrotron radiation, the power radiated by an individual particle can be written down from a generalized form of the Larmor formula, and this general result is reproduced by integrating the emissivity over frequency and solid angle. In this section we show that the corresponding calculations for LAE do not agree in general.  We also discuss a mathematical inconsistency that arises in taking the low-frequency limit.
 
 \subsection{Generalized Larmor formula}

The Larmor formula is derived for emission by an accelerated particle by treating the emission in the rest frame of the particle using the electric dipole approximation. The power radiated is a Lorentz invariant, and by writing the square of the acceleration (which appears in the Larmor formula) in terms of variables in any other inertial frame of interest, one obtains the well-known generalization of the Larmor formula. For highly relativistic particles, this formula implies a power proportional to $\gamma^2$ for acceleration perpendicular to the velocity, as for synchrotron radiation, and a power independent of $\gamma$ for acceleration parallel to the velocity, as for LAE. The instantaneous power radiated for acceleration by an electric field parallel to the velocity is 
\be
P(\chi)={q^4E^2(\chi)\over6\pi\varepsilon_0m^2c^3}.
\label{Larmor1}
\ee
For the triangular wave form of a LAEW, the mean power radiated, averaged over one period of the LAEW, is
\be
{\bar P}={q^2\omega_E^2\over18\pi\varepsilon_0c}
={q^2\Omega^2\gamma_{\rm max}^2\over18\pi\varepsilon_0c|F(\pi/2)|^2},
\label{Larmor2}
\ee
where $\gamma_{\rm max}=(\omega_E/\Omega)|F(\pi/2)|$ is used in the second form. The mean power (\ref{Larmor2}) does not agree with the final form in (\ref{Larmor6}).

\subsection{Inconsistencies}

Although the expressions for the power radiated, (\ref{Larmor5}) or (\ref{Larmor6}),  in LAE has the same functional form as the expression (\ref{Larmor2})  derived from the Larmor formula, the two results differ by a numerical factor. This inconsistency becomes worse when one considers the case of a test charge, $u_0\ne0$. The Larmor formula implies a power (\ref{Larmor2}) independent of $u_0$, whereas (\ref{Larmor6}) implies that the power radiated for $u_0\ne0$ does depend on $u_0$. Moreover, for an arbitrary wave form, the ${\bar P}$ from the Larmor formula involves an average of $T^2(\chi)$ over $\chi$, in the counterpart of (\ref{Larmor2}), whereas (\ref{Larmor6}) depends on the wave form only through $T'(\pi/2)$ and $F(\pi/2)$; these formulae cannot be the same in general.

Another inconsistency arises from an uncritical interpretation of (\ref{Larmor1}) for a LAEW. On the one hand, (\ref{Larmor1}) suggests that the maximum power radiated is when $E^2(\chi)$ is maximum. On the other hand, our detailed calculation shows that the power is strongly concentrated around the phase where the Lorentz factor is maximum, which is the phase where $E^2(\chi)$ vanishes, and the acceleration is instantaneously zero. This is reflected in the power radiated (\ref{Larmor6}) depending on the wave form only at the phases $\chi=\pi/2+n\pi$ where the Lorentz factor has maxima $\gamma_\pm$. 

We note that inconsistencies in the treatment of emission by a linearly accelerated charge have been recognized for over a century, specifically in connection with uniform acceleration \citep{d49}. One outstanding problem centers around the statement ``a uniformly accelerated charge does not radiate.'' The radiation reaction force is proportional to the time derivative of the acceleration, which is zero for uniform acceleration, seemingly supporting the argument that the power radiated is zero. However, this is obviously inconsistent with the Larmor formula, (\ref{Larmor1}). Suppose one attempts to apply the method we use to uniform acceleration. Three difficulties arise: there is no natural frequency $\Omega$, there is no $\gamma_{\rm max}$, and $T(\chi)=$ constant implies $T'(\chi)=0$, so that the cubic term in $\chi'$ in (\ref{phase0}) is zero, and the Airy approximation is invalid. Nevertheless, one can start with a periodic $T(\chi)$ and approach the limit of uniform acceleration, by assuming $T(\chi)\to1$ for $-\pi/2<\chi<\pi/2$, and $T(\chi)\to-1$ for $\pi/2<\chi<3\pi/2$. The emission of LAE is then dominated by the phases, $\chi=\pi/2,3\pi/2$, where the Lorentz factor reaches its extrema, and the electric field passes through zero by abruptly changing sign. Such a treatment of LAE due to uniform acceleration implies that the emission is determined by the way the uniform acceleration is turned on and off, which is one of the ingredients in overcoming the known inconsistencies \citep{d49}.

\subsection{Critique of Larmor formula for LAE}

The foregoing inconsistency suggests that, unlike the case of synchrotron radiation, LAE cannot be treated exactly using the Larmor formula. To understand why this is the case, we need to consider the conditions under which (\ref{Larmor1}) is valid.

A standard derivation of the Larmor formula involves calculating the Poynting vector due to the electric and magnetic fields of the accelerated charge, and integrating over a fixed, large sphere to find the power crossing this sphere. The power lost by the particle is at the retarded time compared with the power escaping. This invalidates any interpretation of the dependence of the Larmor formula (\ref{Larmor1}) on $\chi$ in terms of the time dependence of the power radiated. However, for the average (\ref{Larmor2}) over a periodic motion, the average over the actual time and the retarded time are equivalent. Hence, this argument does not invalidate (\ref{Larmor2}), and so does not explain the inconsistency.

The power escaping from the fixed sphere is equated to the power lost by the particle. This assumption is not valid for LAE in general. The motion of the radiation pattern with the particle inside the fixed sphere implies that the total electromagnetic energy within the sphere is not constant: the radiant energy inside the fixed sphere is changing systematically as a function of time. Hence the assumption that the power radiated balances the power lost by the particle is not valid over any fixed time interval. However, for periodic motion, provided that one considers only the power averaged over a period, and provided that there is no average drift motion, the Larmor formula should be valid. This suggests that the mean power (\ref{Larmor2}) should be correct for a background particle (but not for a test charge). Nevertheless, the expressions for the power do not agree even for a background particle. We conclude that this argument does not explain the inconsistency.

This standard derivation of the Larmor formula also involves assuming that the emission in the instantaneous rest frame can be treated in the electric dipole approximation. For a charge that is instantaneously at rest, undergoing an acceleration, ${\bi a}$, the power in electric dipole radiation is proportional to $|q{\bi a}|^2$. The inconsistency can be resolved if the dipole approximation is not valid. That this is indeed the case for LAE can be seen as follows. As shown by our treatment in Sec.~\ref{sect:emissivity}, the emission of LAE is dominated by the phase where the electric field and the acceleration are instantaneously zero. This is inconsistent with the dipole approximation in the instantaneous rest frame. We conclude that the Larmor formula (\ref{Larmor1}) is not valid for LAE because the assumption that the emission may be treated as electric dipole emission in the instantaneous rest frame in not valid. 

\subsection{Mathematical dilemma}
\label{dilemma}

A mathematical inconsistency arises whenever one attempts to take the low-frequency limit while retaining the $\theta$ dependence. A simple example of this is the low frequency limit of the emissivity (\ref{eta2a}). Assuming $z_c\to0$, one sets the argument of the Airy function to 0 to find
$\eta(\omega,\theta)\propto\Theta\,z_c^2{\rm Ai}^2(0)$, with $\Theta=\gamma^2\theta^2$. In this approximation, the integral over angle diverges, whereas the physically significant contribution is known to come from $\Theta\lapprox1$ ($\theta\lapprox1/\gamma$). Another example where an inconsistency arises is when one attempts to integrate the emissivity (\ref{eta2a}) over angle directly. This may be achieved using the indefinite integral
\be
\int\rmd\Theta\,\Theta\,{\rm Ai}^2(z)={1\over3z_c^2}\left[
z(z-3z_c){\rm Ai}^2(z)-z_c(z-3z_c){\rm Ai}'^2(z)+z_c^2{\rm Ai}(z){\rm Ai}'(z)
\right],
\label{Larmor7}
\ee
with $z=z_c(1+\Theta)$. The validity of (\ref{Larmor7}) is confirmed by differentiating both sides and using the differential equation for Airy functions, ${\rm Ai}''(z)=z{\rm Ai}(z)$. The definite integral over $0<\theta<\infty$ is given by minus the right hand side, with $z\to z_c$. The numerator in the resulting expression, written terms of Macdonald functions using (\ref{Airy2}), gives
\be
2z_c^2{\rm Ai}^2(z_c)-2z_c{\rm Ai}'^2(z_c)-{\rm Ai}(z_c){\rm Ai}'(z_c)=
{3\pi^2\xi_c^2\over2}\left[
K_{1/3}^2(\xi_c)-K_{2/3}^2(\xi_c)+{1\over3\xi}K_{1/3}(\xi_c)K_{2/3}(\xi_c)
\right],
\label{Larmor8a}
\ee
with $\xi_c={2\over3}z_c^{3/2}$. The power per unit frequency becomes
\be
P(\omega)={q^2\Omega\over2\pi^2\varepsilon_0c}\xi_c^2
\left[
K_{1/3}^2(\xi_c)-K_{2/3}^2(\xi_c)+{1\over3\xi_c}K_{1/3}(\xi_c)K_{2/3}(\xi_c)
\right].
\label{Pomega1a}
\ee
On integrating the result (\ref{Pomega1a}) over frequency, using (\ref{schwinger1}) to find $5\pi^2/144$ and $7\pi^2/144$ for the first two integrals, with the third giving $\pi^2/12$, the result (\ref{Larmor5}) is reproduced. The inconsistency arises from the form of the power spectrum at low frequencies. This form is determined by the $\xi_c$-dependence in (\ref{Pomega1a}):
\be
\xi_c^2
\left[
K_{1/3}^2(\xi_c)-K_{2/3}^2(\xi_c)+{1\over3\xi_c}K_{1/3}(\xi_c)K_{2/3}(\xi_c)
\right]\approx
\left\{
\begin{array}{ll}
2\pi/3\sqrt{3}
&\quad\xi_c\ll1,\\
(\pi\xi_c/72)^{1/2}\rme^{-2\xi_c}.
&\quad\xi_c\gg1.
\end{array}
\right.
\label{Pomega2}
\ee
The low-frequency expansion is inconsistent with (\ref{eta7}), and is simply wrong. However, there is no obvious mathematical error. We note that the numerator on the right hand side of (\ref{Larmor7}) may be rewritten
\be
2z_c^2{\rm Ai}^2(z_c)-2z_c{\rm Ai}'^2(z_c)-{\rm Ai}(z_c){\rm Ai}'(z_c)
=4z_c^2{\rm Ai}^2(z_c)-z_c^{1/2}{\rmd\over\rmd z_c}\left[
z_c^{1/2}{\rmd\over\rmd z_c}{\rm Ai}^2(z_c)]
\right],
\label{Larmor7a}
\ee 
that in carrying out the integral over frequency the final term in (\ref{Larmor7a}) integrates to zero, and that the correct form for the power spectrum is obtained simply by ignoring the final term in (\ref{Larmor7a}). However, the final term is not zero and there is no obvious mathematical justification for neglecting it.

We are unable to resolve this inconsistency to our own satisfaction. The following is our opinion on the most plausible source of the inconsistency. At a more fundamental level, the derivation of the power radiated, given by (\ref{Larmor5}) or (\ref{Larmor6}), involves a singular integral. The appearance of a singular integral was noted by \cite{w59} in his derivation of formulae that are now standard for synchrotron radiation. To carry out the integral over angle in the synchrotron case, the procedure used by \cite{w59} involves writing the square of (\ref{Airy1}) in the form
\be
\left|\int_{-\infty}^\infty\rmd\chi\,\rme^{\rmi(a\chi+b^3\chi^3/3)}\right|^2=
\int_{-\infty}^\infty\rmd\chi_1\int_{-\infty}^\infty\rmd\chi_2\,\rme^{\rmi[a(\chi_1-\chi_2)+b^3(\chi_1^3-\chi_2^3)/3]},
\label{Airy12}
\ee
changing the variables to $x=(\chi_1-\chi_2)/2$, $y=(\chi_1+\chi_2)/2$, performing the integral over $y$, and then integrating over angle. This leads to a singular $x$-integral, due to a factor $1/x^2$ in the integrand in the synchrotron case, and to a factor $1/x^{5/2}$ in the case of LAE. Our suggestion is that the singular nature of the integral invalidates taking the low-frequency limit for arbitrary angles of emission. We note that \cite{w59} used partial integration in dealing with the singular integrals in the treatment of synchrotron emission, and that our use of partial integration in (\ref{eta5a}) is an indirect way of avoiding this difficulty.

\section{Discussion}
\label{sect:discussion}

In this section we discuss the properties of LAE and comment on the significance in the application to pulsars.

\subsection{Properties of LAE}

Our results suggest the following interpretation of LAE. As a charge is accelerated, over the first half phase of the LAEW it emits a pulse of radiation in the forward direction of duration $\Delta t_{\rm emit}\sim\pi/\Omega$. The radiation received by a distant observer has a shorter time scale $\Delta t_{\rm rec}=(1-\beta)\Delta t_{\rm emit}$, where $\beta c$ is the speed at which the particle is approaching the observer. The shortest time scale on which structure can be observed in the pulse is $\Delta t_{\rm rec}\approx\Delta t_{\rm emit}/2\gamma_\pm^2\approx\pi/2\Omega\gamma_\pm^2$. Such a pulse of radiation has Fourier components up to a frequency $\omega\sim1/\Delta t_{\rm rec}\approx2\Omega\gamma_\pm^2/\pi$. This radiation is characteristic of emission by a particle with Lorentz factor $\gamma_\pm$ and is confined to a cone of half angle $\sim1/\gamma_\pm$ about the direction of motion. This simple model reproduces the characteristic frequency and angular distribution implied by the emissivity (\ref{eta2a}).

The characteristic maximum frequency is LAE is found to be $\Omega\gamma_\pm^2$, where $\Omega$ is the frequency of the LAEW. This result is derived assuming that the initial Lorentz factor, $\gamma_0$, is much smaller than $\gamma_{\rm max}\sim\omega_E/\Omega$, $\omega_E=|qE_0|/mc$, which is the maximum Lorentz factor that a background particle reaches in the LAEW, with $\gamma_\pm=\gamma_{\rm max}\pm\gamma_0$. In the opposite limit, $\gamma_0\gg\gamma_{\rm max}$, the effect of the LAEW may be treated using perturbation theory, and the characteristic frequency of LAE  in $\Omega\gamma_0^2$ \citep{m78}. In the case where LAE may be treated using perturbation theory, it may be regarded as a form of free-electron maser emission \citep{fk04}. 

\subsection{Application of LAE to pulsars}

Our original motivation for this investigation was the application of LAE in a LAEW to pulsars and magnetars. We identify four possible implications of LAE for pulsars: (a) LAE as a high-energy emission process, (b) LAE as a possible source of secondary pairs, (c) LAE as a damping mechanism for the LAEW, and (d) LAE as a coherent radio emission mechanism. Application to coherent emission is discussed briefly in paper~1, and requires a more detailed discussion than is appropriate here. We comment on each of the other three possible implications.

For LAE to account for observed high-energy emission from pulsars, it must be able to account for the frequency and the power in the observed emission. The characteristic maximum frequency of LAE, $\omega\approx\Omega\gamma_{\rm max}^2$, for the numerical, $\gamma_{\rm max}\sim10^6$--$10^7$ and $\Omega\sim10^6\rm\,s^{-1}$, estimated by \cite{letal05} corresponds to a photon energy of several tens of kilovolts. This suggests that LAE is not a viable emission mechanism for gamma-ray emission. However, before dismissing LAE as a gamma-ray emission we need to consider three effects that allow emission at higher frequencies. First, the estimates $\gamma_{\rm max}\sim10^6$ and $\Omega\sim10^6\rm\,s^{-1}$ might be too conservative for a realistic model; we return to this point below. Second, the frequency $\omega\approx\Omega\gamma_{\rm max}^2$ applies in the frame in which the oscillations in the LAEW are purely temporal, and there is a boost associated with the outward propagation of the LAEW in the pulsar frame. (Both the frequency of the emission and the frequency of the LAEW are transformed and the transformed frequency may be estimated using the invariant, in the notation used in paper~1, $kK$.) For this effect to be large, the phase speed, $\beta_Vc$, must be relatively close to $c$, so that the boost is by $\gamma^*=\beta_V(\beta_V^2-1)^{-1/2}$ is large. We have no reason to expect $\gamma^*$ to be particularly large. Third, LAE might be due to a test charge with $\gamma_0\gg\gamma_{\rm max}$, when LAE may be regarded as a type of free-electron maser emission \citep{fk04}.  Although very high energy `primary' particles appear in some pulsar models, it is not obvious how that would be accelerated in an oscillating model. We conclude that LAE in a LAEW is not a plausible candidate for gamma-ray emission in pulsars.

For the photons associated with LAE to produce pairs, their energy must exceed an MeV. The estimate of a maximum energy of of several tens of kilovolts is sufficiently close to this threshold to require a more detailed discussion. Before considering this, it is relevant to consider the power in LAE. For LAE to be important in generating pairs, the power in LAE must be a significant fraction of the total power involved. It is reasonable to assume that in an oscillating model, the power released (from rotational energy for ordinary pulsars, and from magnetic energy for magnetars) is channeled through LAEWs. The transfer of energy from the LAEW to pairs through LAE may be regarded as a dissipation process for the LAEW. For LAE to be important in generating pairs, not only must its frequency exceed an MeV but also the LAEW must lose a substantial fraction of its energy to LAE before propagating out of the light cylinder.

The damping rate of the LAEW due to LAE may be estimated by multiplying the power in LAE per particle by the number of particles, and dividing by the energy in the LAEW. The power per unit volume follows from the power per background particle, given by (\ref{Larmor5}), times the number density of background particles, ${\bar n}$ say. Ignoring factors of order unity, the damping rate due to LAE is of order $\Gamma\approx\sigma_T\,{\bar n}c$, where $\sigma_T$ is the Thomson cross section. The number density determines the plasma frequency, which is identified as $\Omega\gamma_{\rm max}^{1/2}$ \citep{letal05}. It follows that the damping decrement, $\Gamma/\Omega$, is of order $(r_0\Omega/c)\gamma_{\rm max}$, where $r_0$ is the classical radius of the electron. LAE is important energetically provided that the LAEW remains within the magnetosphere for $\sim\Omega/\Gamma$ wave periods.

\subsection{LAE and secondary pairs}

Secondary pair production in pulsars requires a source of photons with energies $>1\,$MeV. The sources considered in conventional models are curvature emission and resonant Thomson scattering by primary particles. In an oscillating model LAE is an additional possibility. For LAE to be viable as the source of secondary pairs, two conditions need to be satisfied: the photon energy must exceed an MeV, and the power in LAE must be sufficient to account for the required number of pairs.

Consider a model in which there is a large number of localized, transient LAEWs in the polar cap region, with the pairs in the LAEW created through LAE. Let the number density of pairs be a multiplicity, $M$, times the Goldreich-Julian density, so that the frequency of the LAEW is $\omega_{\rm max}\sim (M\Omega_r\Omega_c)^{1/2}\gamma_{\rm max}^{3/2}$, where $\Omega_r=2\pi/P$ is the rotation frequency of a pulsar with period $P$, and $\Omega_c=(mc^2/\hbar)(B/B_c)$ is the cyclotron frequency, with $B_c=4.4\times10^9\rm\,T$ the Schwinger field. The threshold condition, $\omega_{\rm max}>2mc^2/\hbar$, requires 
\be
{M\over P}\,{B\over B_c}\gamma_{\rm max}^3\gapprox10^{20},
\label{pair1}
\ee
where $P$ is in seconds. The fraction of the energy lost by a LAEW as it propagates outward through the pulsar magnetosphere can be estimated by multiplying this damping decrement by the number of oscillations before the LAEW leaves the magnetosphere. Assuming propagation at close to the speed of light this number is of order $\Omega/\Omega_r$. Hence, the fraction of the energy lost to LAE is of order $(r_0\Omega^2/\Omega_rc)\gamma_{\rm max}\sim Mr_0\Omega_c/c=M\alpha(B/B_c)$, where $\alpha\approx1/137$ is the fine structure constant. We conclude that LAE is energetically important in an oscillating model for a pulsar provided the condition
\be
M\,{B\over B_c}\gapprox10^{2}
\label{pair2}
\ee
is satisfied. No extreme values are required to satisfy (\ref{pair1}) for all pulsars, and although (\ref{pair2}) is satisfied for $M\gapprox10^3$ for ordinary pulsars, it requires a rather extreme multiplicity for recycled (millisecond) pulsars. We conclude that an oscillating model with secondary pair production due to LAE is consistent with parameters otherwise regarded as plausible.

\section{Conclusions}
\label{sect:conclusions}

Our main objective in this paper is to develop the theory of LAE to see if it is viable as an emission process in an oscillating model for pulsars, with the oscillations described in terms of a large amplitude electrostatic wave (LAEW). We find the following properties of LAE in a LAEW; these properties apply in the inertial frame in which the oscillations are purely temporal:
\begin{enumerate}
\item The emission is dominated by the phase of the LAEW where the electric field passes through zero and the particles have their maximum Lorentz factor, $\gamma_\pm=\gamma_{\rm max}\pm u_0$, where  $\gamma_{\rm max}\approx|q|E_0/mc\Omega$ is determined by the frequency, $\Omega$, of the LAEW and its amplitude, $E_0$, and where $u_0=\gamma_0\beta_0$ is the 4-velocity of the particle at the phase where the electric field is equal to $\pm E_0$, depending on the sign of the charge.
\item The characteristic maximum frequency of LAE is $\Omega\gamma_\pm^2$ for $\gamma_0\ll\gamma_{\rm max}$. (We do not consider the case $\gamma_0\gg\gamma_{\rm max}$ when the emission may be regarded as a form of free electron maser emission.)
\item LAE is emitted in the forward direction in one half period, as the particle propagates in the forward direction, and in the backward direction in the other half period; it is concentrated on the surface of a cone of half angle $\approx1/\gamma_\pm$ about this direction, and is zero at at the center of the cone.
\item The power per unit frequency increases $\propto\omega^{4/3}$ at $\omega\ll\Omega\gamma_\pm^2$.
\item The total power radiated is given by (\ref{Larmor6}). The total power is not given correctly by the generalized Larmor formula (\ref{Larmor1}), or its average (\ref{Larmor2}) over a period of oscillation; we argue that this is because the dipole approximation is not valid in the instantaneous rest frame.
\item A mathematical inconsistency arises in taking the low-frequency limit of the emissivity at an arbitrary angle of propagation. We attribute this inconsistency to an underlying singular integral, but have been unable to reformulate the theory so that this inconsistency does not arise.
\end{enumerate}

We discuss the application of LAE to pulsars, both as a high-energy emission process and as a maser emission process at radio frequencies. Our conclusions are:
\begin{enumerate}
\item  LAE is implausible as a gamma-ray emission, at photon energies $\gg1\,$MeV. 
\item It seems possible that LAE-photons could lead to secondary pair production in an oscillating model. 
\item Maser LAE is a possible coherent radio emission mechanism only for relatively small amplitude
LAEWs (paper~1).
\end{enumerate}

\acknowledgements

We thank Matthew Verdon for helpful comments on the manuscript.

\end{document}